# Graph-theoretical Constructions for Graph Entropy and Network Coding Based Communications

Maximilien Gadouleau, *Member, IEEE* and Søren Riis, *Member, IEEE*

**Abstract**

The guessing number of a directed graph (digraph), equivalent to the entropy of that digraph, was introduced as a direct criterion on the solvability of a network coding instance. This paper makes two contributions on the guessing number. First, we introduce an undirected graph on all possible configurations of the digraph, referred to as the guessing graph, which encapsulates the essence of dependence amongst configurations. We prove that the guessing number of a digraph is equal to the logarithm of the independence number of its guessing graph. Therefore, network coding solvability is no more a problem on the operations made by each node, but is simplified into a problem on the messages that can transit through the network. By studying the guessing graph of a given digraph, and how to combine digraphs or alphabets, we are thus able to derive bounds on the guessing number of digraphs. Second, we construct specific digraphs with high guessing numbers, yielding network coding instances where a large amount of information can transit. We first propose a construction of digraphs with finite parameters based on cyclic codes, with guessing number equal to the degree of the generator polynomial. We then construct an infinite class of digraphs with arbitrary girth for which the ratio between the linear guessing number and the number of vertices tends to one, despite these digraphs being arbitrarily sparse. These constructions yield solvable network coding instances with a relatively small number of intermediate nodes for which the node operations are known and linear, although these instances are sparse and the sources are arbitrarily far from their corresponding sinks.

The authors are with the School of Electronic Engineering and Computer Science, Queen Mary, University of London, London, E1 4NS, UK (e-mail: {mgadouleau, smriis}@eecs.qmul.ac.uk).

# I. INTRODUCTION

Network coding [1] is a protocol which outperforms routing for multicast networks by letting the intermediate nodes manipulate the packets they receive. In particular, linear network coding [2] is optimal in the case of one source; however, it is not the case for multiple sources [3], [4]. Although for large dynamic networks, good heuristics such as random linear network coding [5], [6] can be used, for a given static network maximizing the amount of information that can be transmitted is fundamental. Solving this problem by brute force, i.e. considering all possible operations at all nodes, is computationally prohibitive. In this paper, we reduce this problem to finding a maximum independent set in an undirected graph determined by the network coding instance.

Network coding also opens many new questions about network design (see [7], [8] for examples of networks with interesting properties). Clearly, dense graphs with a large number of edges between the nodes can transmit a large amount of information; similarly, a small diameter is a good property for information transfer; finally, a large number of intermediate nodes between the sources and the sinks is preferable. However, in this paper, we introduce classes of networks that are arbitrarily sparse, with arbitrarily high diameters, and with a relatively small number of intermediate nodes, yet on which all the requested information can be transmitted. Furthermore, for these graphs, the demands of the sinks can be satisfied over any alphabet, and linear combinations are sufficient. Therefore, our work provides different guidelines on the design of networks which take advantage of network coding. The results in this paper are based on the study of the guessing number of digraphs, reviewed below.

The guessing number of digraphs is a concept introduced in [9], which connects graph theory, network coding, and circuit complexity theory. In [9] it was proved that an instance of network coding with $n$ sources and $n$ sinks on an acyclic network (referred to as a multiple unicast network) is solvable over a given alphabet if and only if the guessing number of a related digraph is equal to $n$. Moreover, it is proved in [9], [10] that any network coding instance can be reduced into a multiple unicast network. Therefore, the guessing number is a direct criterion on the solvability of network coding. Similarly, the linear guessing number evaluates the solvability of a network coding instance by using linear combinations only. By determining these two quantities, the performance of linear network coding can then be compared to that of general network coding. In [11], the guessing number is also used to disprove a long-standing open conjecture on circuit complexity. In [12], the guessing number and linear guessing number of digraphs were studied, and bounds on the guessing number of some particular digraphs were derived.

The guessing number is equal to the entropy of the same digraph [11], thus tying this quantity with fundamental problems of information theory. For instance, by relying heavily on [13], [14] and [15], it was shown that the entropy of a digraph might not be determined by the use of Shannon inequalities alone [16]. Similarly, the information defect is related to the so-called public entropy [16]. We would like to emphasize that the graph entropy for digraphs considered in this paper is fundamentally different to the graph entropy for undirected graph introduced by Körner in [17] (see [18] for a review of that quantity).

Let us give a brief description of the guessing game with $n$ players, viewed as vertices on a digraph $D$, and an alphabet of size $s$. All the players are assigned an element of the alphabet (collectively referred to as a configuration), and each player knows the values assigned to all the players in its in-neighborhood. It does not, however, know its own value, and the goal of the game is to guess it correctly. Clearly, the values cannot all be guessed correctly every time. If the players do not collaborate, the probability that all guesses are correct is exactly $s^{-n}$. However, the players may elaborate a collaborative strategy (referred to as a protocol) which increases the probability of success. For instance, suppose we play the game on the clique $K_n$, where each player knows the values assigned to all the other vertices. A common strategy could be the following: each player guesses the opposite of the sum (modulo $s$) of all the values it sees. Any configuration whose sum modulo $s$ is zero will be correctly guessed, hence raising the success probability to $s^{-1} = s^{(n-1)-n}$ (this is, in fact, optimal). The guessing number is then defined as the maximum over all protocols of the gain from the trivial guessing strategy. For instance, the guessing number of the clique on $n$ vertices is $n - 1$.

Suppose now the players have a helper, whose aim is to make all players guess correctly every time. This helper is limited: he or she can only send the same information to all the players. The information defect is defined to be the minimum amount of information the helper must send, and it is strongly connected to the guessing number. For instance, in $K_n$, the players will be able to infer their own value if the helper sends them the sum of all values modulo $s$. Only one symbol of information is required, therefore the information defect of the clique on $n$ vertices is equal to $1$. While the guessing number $g(D, s)$ represents the amount of information that can be guessed by the players, the information defect $b(D, s)$ is the amount of common information the players need to guess correctly. The information defect is shown in [8] to be equal to the length of a minimal index code induced on the graph $D$ (see [19] for more on index coding and its relation to network coding).

This paper has two main contributions. First, we introduce a graph on all the possible configurations of a digraph, referred to as the *guessing graph*, which encapsulates the dependencies amongst fixed configurations of the same protocol. We then show that the guessing number of a digraph is equal to the logarithm of the independence number of its guessing graph. The study of the guessing graph then yields the following results.

- Solvability of network coding is no more a problem of determining the appropriate operations at each intermediate node. It is now turned into a problem on the possible messages that could be transmitted through the network by using network coding, and the operations which transmit these messages can then be easily determined. This simplification significantly reduces the search space, which only depends on the number of nodes in the graph and on the alphabet size.
- The problem of solvability of network coding is reduced to a decision problem on the independence number of undirected graphs. Although the guessing graph has an exponential number of vertices, it has a large automorphism group, which could be taken advantage of. We show that finding maximum independent sets on this graph is actually a problem closely related to the design of error-correcting codes. This parallels the results in [20], where it was shown that some classes of network coding instances are solvable if and only if codes with certain parameters exist.
- Using graph theoretic results, we are then able to provide chains of bounds on the guessing number of a digraph based on the properties of its guessing graph. For instance, we obtain that for large enough alphabets, the guessing number is at least equal to the minimum in-degree of a vertex in the digraph, and the fixed configurations attaining this bound form an MDS code.
- The relationship between the guessing game and public information (or equivalently, between public and private entropy) unveiled in [11] is clarified, as we show that the information defect is equal to the chromatic number of the guessing graph. This enables us to prove that these problems are asymptotically equivalent.
- The guessing graph is extremely well-behaved when digraphs are combined. We exhibit some types of digraph union which do not increase the ratio between the guessing number and the number of vertices in the digraph. Also, the guessing graph illustrates the relationships between the guessing numbers of the same digraph over different alphabets. We prove that playing the guessing game on a digraph over an extension field is equivalent to playing the guessing game on several copies of the same digraph linked to one another over the base field.

We would like to emphasize the fundamental difference between our work and the literature where conflicts in networks were represented as adjacent vertices in graphs [21]–[23]. In the literature, the vertices of the different graphs and hypergraphs previously proposed are routes or links amongst nodes or coding functions instead of messages or configurations. Therefore, these do not convert the network coding problem into a problem on messages. Indeed, the vertices of the so-called "link graph" in [21] are the routes from the inputs to the outputs, and two routes conflict if they intersect. Also, the vertices correspond to the cumulative coding functions at each node in [22], and the conflicts amongst functions are represented via a hypergraph. Moreover, the vertices of the so-called "conflict graph" in [23] represent a node in the network along with part of its out-neighbors.

The second contribution is the construction of specific digraphs with high linear guessing numbers, thus yielding solvable network coding instances.

- For a finite number $n$ of source-sink pairs, we introduce a construction of digraphs based on cyclic codes, thus tying another link between network coding and error-correcting codes. All the information about the digraph, and especially its guessing number, are available from the generator polynomial of the code. In particular, the class of digraphs generated by the simplex codes produce network coding instances with bottlenecks on the order of $\log n$ only.

- For unbounded parameters, we determine a way of combining two digraphs, referred to as the strong product, which takes full advantage of the structure of the two original digraphs in order to yield a high guessing number. Using this technique, we construct network coding instances as sparse as possible in terms of edges provided the number of edges tends to infinity, where the shortest path between a source and the corresponding sink is arbitrarily long, and where the number of intermediate nodes is small compared to the number of sources. These instances are solvable over any alphabet and linearly solvable over any field.

The rest of the paper is organized as follows. Section II reviews some necessary background on graph theory, guessing games, and error-correcting codes. Section III introduces and investigates the properties of the guessing graph. In Section IV, we introduce a class of digraphs based on cyclic codes for which we determine the binary linear guessing number. Section V studies the maximum guessing number of digraphs and introduces families of graphs with asymptotically highest guessing numbers. Finally, Section VI provides some comments and presents some open problems.

## II. Preliminaries

### A. Graphs and digraphs

An independent set in a graph is a set of vertices where any two vertices are non-adjacent. The *independence number* $\alpha(G)$ of an undirected graph $G$ is the maximum cardinality of an independent set.

We also denote the maximum degree and the clique, chromatic, and fractional chromatic numbers of an undirected graph $G$ as $\Delta(G)$, $\omega(G)$, $\chi(G)$, and $\chi^*(G)$, respectively (see [24] for definitions of these parameters). For a connected vertex-transitive graph which is neither an odd cycle nor a complete graph, we have [24, Corollary 7.5.2]

$$\omega(G) \leq \chi^*(G) = \frac{|V(G)|}{\alpha(G)} \leq \chi(G) \leq \Delta(G).$$

Also, it was shown in [25] that for a non-complete $\kappa$-connected graph on $n$ vertices which is regular with degree $d$, the independence number is lower bounded by

$$\alpha(G) \geq \frac{n(d+1)}{\kappa}\left\{1 - \sqrt{1 - 2\frac{\kappa}{(d+1)^2}}\right\} \geq \frac{n}{d+1}. \tag{1}$$

The chromatic number and the independence number of a vertex-transitive graph are related by [26] (using the no-homomorphism lemma in [27])

$$\chi(G) \leq (1 + \log \alpha(G)) \max_{H \text{ induced}} \frac{|V(H)|}{\alpha(H)} = (1 + \log \alpha(G))\frac{|V(G)|}{\alpha(G)}. \tag{2}$$

We now review four types of products of graphs; all products of two graphs $G_1$ and $G_2$ have $V(G_1) \times V(G_2)$ as vertex set. We denote tow adjacent vertices $u$ and $v$ in a graph as $u \sim v$.

- First, in the *co-normal product* $G_1 \oplus G_2$, we have $(u_1, u_2) \sim (v_1, v_2)$ if and only if $u_1 \sim v_1$ or $u_2 \sim v_2$. We have

$$\alpha(G_1 \oplus G_2) = \alpha(G_1)\alpha(G_2). \tag{3}$$

- Second, in the *lexicographic product* (also called composition) $G_1 \cdot G_2$, we have $(u_1, u_2) \sim (v_1, v_2)$ if and only if either $u_1 = v_1$ and $u_2 \sim v_2$, or $u_1 \sim v_1$. Although this product is not commutative, we have

$$\alpha(G_1 \cdot G_2) = \alpha(G_1)\alpha(G_2).$$

- Third, in the *strong product* $G_1 \boxtimes G_2$, we have $(u_1, u_2) \sim (v_1, v_2)$ if and only if either $u_1 = v_1$ and $u_2 \sim v_2$, or $u_2 = v_2$ and $u_1 \sim v_1$, or $u_1 \sim v_1$ and $u_2 \sim v_2$.

- Fourth, in the *cartesian product* $G_1 \square G_2$, we have $(u_1, u_2) \sim (v_1, v_2)$ if and only if either $u_1 = v_1$ and $u_2 \sim v_2$, or $u_2 = v_2$ and $u_1 \sim v_1$. We have

$$\chi(G_1 \square G_2) = \max\{\chi(G_1), \chi(G_2)\},$$
$$\alpha(G_1 \square G_2) \leq \min\{\alpha(G_1)|V(G_2)|, \alpha(G_2)|V(G_1)|\}.$$

Throughout this paper, we shall only consider *simple* digraphs, which have no loops and no repeated edges. However, we do allow edges in both directions between two vertices, referred to as *bidirectional* *edges* (we shall abuse notations and identify a bidirectional edge with a corresponding undirected edge). In other words, the digraphs considered here are of the form $D = (V, E)$, where $E \subseteq V^2 \setminus \{(v, v) : v \in V\}$. We shall denote the number of vertices of the digraph as $n$ unless otherwise specified. The adjacency matrix $\mathbf{A}_D$ of a digraph $D$ on $n$ vertices is the $n \times n$ binary matrix such that $a_{i,j} = 1$ if and only if $(v_i, v_j) \in E(D)$. For any vertex $v_i$ of $D$, its in-neighborhood, denoted as $N_-(v_i)$, is the set of all vertices $v_j$ such that $(v_j, v_i) \in E(D)$, and its in-degree is the size of its in-neighborhood. We say that a digraph is *strong* if there is a path from any vertex to any other vertex of the digraph. An *independent set* of vertices in a digraph is a set such that no vertex is in the in-neighborhood of another.

The *girth* of a digraph is the minimum length of a directed cycle (we consider a bidirectional edge as a cycle of length 2). A digraph is *acyclic* if it has no directed cycles. In this case, there is an order of the vertices $v_0, v_1, \ldots, v_{n-1}$, referred to as the *topological order*, for which $(v_i, v_j) \in E(D)$ only if $i < j$ (in particular, $v_0$ has in-degree 0). The cardinality of a maximum induced acyclic subgraph of the digraph $D$ is denoted as $mas(D)$. It can be easily shown that $mas(D) \geq \frac{n}{\Delta+1}$, where $\Delta$ is the maximum in-degree of a vertex in $D$.

## B. Guessing game and guessing number

We denote the ring $\mathbb{Z}(s) = \{0, 1, \ldots, s-1\}$ or the field $\mathrm{GF}(s)$ if $s$ is the power of a prime as $[s]$. A *configuration* on a digraph $D$ is a map from its vertex set $V(D)$ to $[s]$, which we shall identify with its image $x = (x_0, x_1, \ldots, x_{n-1})$. A *protocol* $\mathcal{P}$ on $D$ is a mapping between its configurations such that $\mathcal{P}(x)$ is locally defined, i.e. $\mathcal{P}(x)_v = f_v(x_{v_0}, x_{v_1}, \ldots, x_{v_{k-1}})$, where $k = |N_-(v)|$ and $v_i \in N_-(v)$ for all $i$. For any $J \subseteq \{0, 1, \ldots, n-1\}$, we refer to the word $(x_{j_0}, x_{j_1}, \ldots, x_{j_{|J|-1}})$ where the $j_i$s are sorted in increasing order and are all in $J$ as $x_J$. Using this notation, we have $\mathcal{P}(x)_v = f_v(x_{N_-(v)})$. The fixed configurations of $\mathcal{P}$ are all the configurations $x \in [s]^n$ such that $\mathcal{P}(x) = x$. The *guessing number* of $D$ is then defined as the logarithm of the maximum number of configurations fixed by a protocol of $D$:

$$g(D, s) = \max_{\mathcal{P}} \{\log_s |\mathrm{Fix}(\mathcal{P})|\}.$$

This definition actually depends on $s$, and we can also consider the general guessing number $g(D) = \sup_s g(D, s)$.

A protocol is said to be linear if the local functions are linear: $f_v(x_{N_-(v)}) = y_v \cdot x_{N_-(v)}$ for some $y_v \in \mathrm{GF}(s)^{|N_-(v)|}$. The fixed configurations of a linear protocol form a linear subspace of $\mathrm{GF}(s)^n$. The *linear guessing number* of $D$ is the maximum dimension of the set of fixed configurations of a linear protocol of $D$: $g_{\text{linear}}(D, s) = \max_{\mathcal{P} \text{ linear}} \{\dim \text{Fix}(\mathcal{P})\}$. It is shown in [12, Theorem 4.3] that the linear guessing number is given by

$$g_{\text{linear}}(D, s) = n - \min_{\mathbf{A} \in \text{GF}(s)^{n \times n}, \mathbf{A} \leq \mathbf{A}_D} \{\text{rk}(\mathbf{I}_n + \mathbf{A})\}, \tag{4}$$

where $\mathbf{A} \leq \mathbf{B}$ if and only if $a_{i,j} \neq 0$ implies $b_{i,j} \neq 0$. Clearly, we have $g_{\text{linear}}(D, s) \leq g(D, s)$ for all digraphs $D$.

A *set of public messages* $\mathcal{M}$ is a is a partition of the set of configurations into $b$ pieces of the form $\text{Fix}(\mathcal{P}_i)$, i.e. $\bigcup_{0 \leq i \leq b-1} \text{Fix}(\mathcal{P}_i) = [s]^n$. The *information defect* of the digraph $D$ is defined as the logarithm of the minimum cardinality of a set of public messages, and is denoted as $b(D, s) = \min_{\mathcal{M}} \{\log_s |\mathcal{M}|\}$. It was shown in [11] that for any digraph $D$ on $n$ vertices and any $s$, $b(D, s) + g(D, s) \geq n$. We also consider the general information defect $b(D) = \inf_s b(D, s)$.

## C. Relation between guessing games and network coding

We now review how to convert a multiple unicast problem in network coding to a guessing game. Note that any network coding instance can be converted into a multiple unicast without any loss of generality [10], [11]. Let $N$ be an acyclic network with $n$ sources, $n$ sinks, and some intermediate nodes. We suppose that each sink requests an element from an alphabet $[s]$ from a corresponding source. This network coding instance is *solvable* over $[s]$ if all the demands of the sinks can be satisfied at the same time. We assume the network instance is given in its *circuit representation*, where each vertex represents a distinct coding function and hence the same message flows every edge coming out of the same vertex. This circuit representation has $n$ source nodes, $n$ sink nodes, and $m$ intermediate nodes. By merging each source with its corresponding sink node into one vertex, we form the digraph $D_N$ on $m + n$ vertices. In general, we have $g(D_N, s) \leq n$ for all $s$ and the original network coding instance is solvable over $[s]$ if and only if $g(D_N, s) = n$ [11]. Similarly, we have $b(D_N, s) \geq m$ and the instance is solvable if and only if $b(D_N, s) = m$ [11].

Therefore, while network coding considers how the information flows from sources to sinks, the guessing game captures the intuitive notion of how much information circulates through the digraph. A protocol for the guessing game is equivalent to the network coding operations in the original instance. Since all network coding instances can be turned into a guessing game, the guessing game is a fundamental problem in information transit in networks. Conversely, if a digraph $D$ on $m + n$ vertices has an acyclic induced subgraph $M$ of size $m$, then the $n$ vertices outside $M$ can be split in two to form the circuit representation of a network coding instance with $n$ sources, $n$ sinks, and $m$ intermediate nodes.

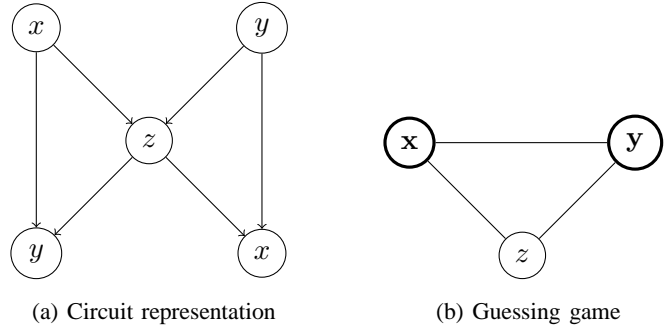

(a) Circuit representation    (b) Guessing game

Fig. 1.  The butterfly network as a guessing game.

We illustrate the conversion of a network coding instance to a guessing game for the famous butterfly network in Figure 1 below. We shall show the vertices corresponding to the source-sink pairs in bold with thick contours henceforth. It is well-known that the butterfly network is solvable over all alphabets (by adding the two incoming messages modulo $s$ in $z$), and conversely it was shown that the clique $K_3$ has guessing number 2 over any alphabet (and the protocol is simple: all nodes guess minus the sum modulo $s$ of their incoming elements).

## D. Error-correcting codes

The weight of a word $x$ in $[s]^n$ is the number of nonzero symbols of $x$ and is denoted as $w(x)$. A code of length $n$ over $[s]$ with minimum Hamming distance $d$ is a set of words in $[s]^n$ such that any two words differ in at least $d$ positions. We denote the maximum cardinality of such a code as $A_s(n,d)$. The Singleton bound asserts that $A_s(n,d) \le s^{n-d+1}$, and this bound is achieved by Maximum Distance Separable (MDS) codes. MDS codes are known to exist for $d \in \{1, 2, n\}$ or when $s$ is the power of prime and satisfies either $s \ge n-1$ or $s = 2^m$, $n = 2^m + 2$, $d \in \{4, n-2\}$ [28, Chapter 11, Section 7].

A binary $(n,k)$ linear code $C$ is a linear subspace of $\mathrm{GF}(2)^n$ with dimension $k$. If $C$ is the row span of a matrix $\mathbf{G} \in \mathrm{GF}(2)^{k \times n}$, we say that $\mathbf{G}$ is a *generator matrix* of $C$. Moreover, if $C$ is the row space of a matrix $\mathbf{G}' \in \mathrm{GF}(2)^{n \times n}$ of rank $k$, we say that $\mathbf{G}'$ is an extended generator matrix of $C$. Alternatively, if $C$ is the dual space of the row space of a matrix $\mathbf{H} \in \mathrm{GF}(2)^{(n-k) \times n}$ (resp., $\mathbf{H}' \in \mathrm{GF}(2)^{n \times n}$ with rank $n-k$), we say that $\mathbf{H}$ is a *parity-check matrix* (resp., extended parity-check matrix) of $C$. By definition, we have $\mathbf{c}\mathbf{H}'^T = \mathbf{0}$ for all $\mathbf{c} \in C$.

A (binary) *cyclic code* is a linear binary code where all the cyclic shifts of a codeword are also codewords. To any vector $c = (c_0, c_1, \ldots, c_{n-1}) \in \mathrm{GF}(2)^n$, we associate the polynomial $c(x) = \sum_{i=0}^{n-1} c_i x^i$.

A cyclic code can then be viewed as an ideal in the ring of polynomials modulo $x^n + 1$, where $n$ is the length of the code. Therefore, a cyclic code is composed of all the multiples of a *generator polynomial* $g(x)$ of degree $n - k$, where $k$ is the dimension of the code. A generator matrix for the code is hence given by $k$ shifts of $g(x)$. Remark that a polynomial generates a cyclic code of length $n$ if and only if it divides $x^n + 1$.

A *constant-weight code* is a binary code consisting of codewords with the same Hamming weight. They have attracted a large interest; a thorough survey is provided in [29], and various upper bounds are derived or reviewed in [30]. The maximum cardinality of a constant-weight code of length $n$, weight $w$, and minimum distance $2d$ (as it is always even) is upper bounded by $\binom{n}{w-d+1}/\binom{w}{w-d+1}$ [31].

### III. THE GUESSING GRAPH OF A DIGRAPH

*A. Guessing graph, guessing number, and information defect*

In this section, we introduce an undirected graph on all possible configurations of a digraph, where an independent set corresponds to a set of fixed configurations of a protocol. As a result, the guessing number of the digraph is equivalent to the logarithm of the independence number of the associated graph.

*Definition 1 (Guessing graph of a digraph):* For any digraph $D$ on $n$ vertices and any integer $s \geq 2$, the $s$-guessing graph of $D$, denoted as $\mathrm{G}(D, s)$, has $[s]^n$ as vertex set and two configurations are adjacent if and only if there is no protocol for $D$ which fixes them both.

Proposition 1 below enumerates some properties of the guessing graph. In particular, Property provides a concrete and elementary description of the edge set which makes adjacency between two configurations easily decidable.

*Proposition 1:* The guessing graph $\mathrm{G}(D, s)$ of a digraph $D$ on $n$ vertices satisfies the following properties:

1) It has $s^n$ vertices.
2) Its edge set is $E = \bigcup_{i=0}^{n-1} E_i(s)$, where $E_i(s) = \{\{x, y\} : x_{N_-(v_i)} = y_{N_-(v_i)}, x_i \neq y_i\}$.
3) It is regular with degree
$$d(\mathrm{G}(D, s)) = \sum_{I \text{ independent}} (-1)^{|I|-1}(s-1)^{|I|} s^{n-|N_-(I)|-|I|},$$
   where $N_-(I)$ is the union of all the in-neighborhoods of vertices in $I$.
4) It is vertex-transitive. More particularly, for any adjacent configurations $x = (x_0, x_1, \ldots, x_{n-1}), y = (y_0, y_1, \ldots, y_{n-1}) \in [s]^n$, we have

- $x + e \sim y + e$ for any $e \in [s]^n$;
- $\pi(x) \sim \pi(y)$ for any $\pi \in \text{Aut}(D)$;
- if $s$ is the power of a prime, $(\lambda_0 x_0, \lambda_1 x_1, \ldots, \lambda_{n-1} x_{n-1}) \sim (\lambda_0 y_0, \lambda_1 y_1, \ldots, \lambda_{n-1} y_{n-1})$ for any family of nonzero scalars $\lambda_i \in \text{GF}(s)$.

*Proof:* Property 1 follows Definition 1. Let us prove Property 2. Let $x, y \in E_i(s)$ for some $i$ and let a protocol with local functions $f_{v_i}$ fix $x$. Then $f_{v_i}(y_{N_-(v_i)}) = f_{v_i}(x_{N_-(v_i)}) = x_i \ne y_i$, hence $\mathcal{P}$ does not fix $y$. Conversely, if $x, y \notin E$ then any protocol satisfying $f_{v_i}(x_{N_-(v_i)}) = x_i$ and $f_{v_i}(y_{N_-(v_i)}) = y_i$ for all $i$ fixes both $x$ and $y$.

Property 4 follows this observation: $x \sim y$ if and only if $(x-y)_{N_-(v_i)} = 0$ and $(x-y)_i \ne 0$ for some $i$. Since the guessing graph is vertex-transitive it is regular and hence we determine the number of edges adjacent to the all-zero configuration $0$. By the inclusion-exclusion principle, we have

$$d(\text{G}(D,s)) = |E \cap \{0\}| = \left| \bigcup_{i=0}^{n-1} E_i(s) \cap \{0\} \right| = \sum_{R \subseteq V} (-1)^{|R|-1} |E_R \cap \{0\}|,$$

where $E_R = \bigcap_{v_i \in R} E_i$, and hence we only have to determine $|E_R \cap \{0\}|$ for all $R \subseteq V$. The configurations $y$ adjacent to $0$ satisfy $w(y_R) = |R|$ and $y_{N_-(R)} = 0$, while $y_{V-N_-(R)-R}$ is arbitrary. If $R$ is not independent, $R \cap N_-(R) \ne \emptyset$ and the two conditions are contradictory; otherwise $R \cap N_-(R) = \emptyset$ and there are $(s-1)^{|R|} s^{n-|N_-(R)|-|R|}$ choices for $y$. ∎

The guessing graph of some particular digraphs can be characterized.

*Example 1:* The following guessing graphs are easy to determine.

- The guessing graph of an acyclic digraph is the complete graph.
- The guessing graph of the clique $K_n$ is given by the Hamming graph $H(s,n)$, where two configurations are adjacent if and only if they are at Hamming distance $1$.
- In the guessing graph of the directed cycle $C_n$, two configurations are adjacent if and only if they are at Hamming distance at most $n-1$.

*Proof:* If $D$ is acyclic, let us sort the vertices in topological order, so that $N_-(v_i) \subseteq \{v_0, v_1, \ldots, v_{i-1}\}$. Consider two distinct configurations $x, y \in [s]^n$, and let $l = \min\{i : x_i \ne y_i\}$, then $x_{N_-(v_l)} = y_{N_-(v_l)}$ and $\{x, y\} \in E_l(s)$.

We now determine the guessing graph of the clique $K_n$. We have $E_i(s) = \{\{x,y\} : x_i \ne y_i, x_{V-\{i\}} = y_{V-\{i\}}\}$ and hence $x$ and $y$ are adjacent if and only if they differ in exactly one coordinate.

We now consider the cycle $C_n$, whose edge set is given by $\{(v_i, v_{i+1 \mod n}) : 0 \le i \le n-1\}$. Suppose $x$ and $y$ are distinct and non-adjacent, then there exists $i$ such that $x_i \ne y_i$. Since $\{x,y\} \notin E_i(s)$, we have $x_{i-1} \neq y_{i-1}$. Applying this recursively, we obtain that all coordinates of $x$ and $y$ must be distinct. Conversely, if $x_i \neq y_i$ for all $i$, then it is clear that $x$ and $y$ are not adjacent. ∎

Clearly, a set of fixed configurations of some protocol forms an independent set in the guessing graph. Theorem 1 below asserts the converse: any independent set can be fixed by some protocol and hence can be viewed as a set of possible transmitted messages on the original network.

*Theorem 1:* A set of configurations in $[s]^n$ are fixed configurations of some protocol for $D$ if and only if they correspond to an independent set in the graph $\mathrm{G}(D,s)$, and hence

$$g(D,s) = \log_s \alpha(\mathrm{G}(D,s)).$$

Moreover, a set of configurations in $[s]^n$ are a set of public messages if and only if it forms a coloring of the guessing graph $\mathrm{G}(D,s)$, and hence

$$b(D,s) = \log_s \chi(\mathrm{G}(D,s)).$$

*Proof:* By definition, any set of fixed configurations of some protocol form an independent set in the guessing graph. Conversely, if $\{x^a\}_{a=0}^{k-1}$ is an independent set of the guessing graph, we shall construct a protocol $\mathcal{P}$ which fixes all $x^a$ configurations. For $0 \leq i \leq n-1$, we define the local functions $\mathcal{P}(x)_{v_i} = f_{v_i}(x_{N_-(v_i)})$ as follows: $f_{v_i}(x^a_{N_-(v_i)}) = x^a_i$ and $f_{v_i}(y) = 0$ if there is no $a$ such that $y = x^a_{N_-(v_i)}$. Note that this is a non-ambiguous assignment, as either $x^a_{N_-(v_i)} \neq x^b_{N_-(v_i)}$ (and the assignments are independent) or $x^a_{N_-(v_i)} = x^b_{N_-(v_i)}$ and $x^a_i = x^b_i$ (the same assignment) for all $a, b \in \{0, 1, \ldots, k-1\}$.

Finally, since a set of public messages is a partition of $[s]^n$ into sets of fixed configurations, it is equivalent to a coloring of the guessing graph. ∎

The guessing numbers of the digraphs mentioned in Example 1 were already determined in [11] or [12]. However, the proof becomes straightforward using Theorem 1.

*Example 2:* If $D$ is acyclic, then $g(D,s) = 0$ and $b(D,s) = n$ for all $s$. This can be intuitively explained as follows: since the digraph has no cycle, no information can circulate around it. Also, the clique satisfies $g(K_n, s) = n-1$, $b(K_n, s) = 1$, which means that the $n-1$ symbols of information received by any vertex can circulate around the digraph. Finally, for the directed cycle we have $g(C_n, s) = 1$, $b(C_n, s) = n-1$, since one symbol of information naturally circulates along the cycle.

In order to illustrate the relevance of this result to network coding, we return to the butterfly network example given in Figure 1. We already showed that it was equivalent to a guessing game on the clique $K_3$. Its binary guessing graph, given by the cube $H(2,3)$, is illustrated in Figure 2. Throughout this paper,

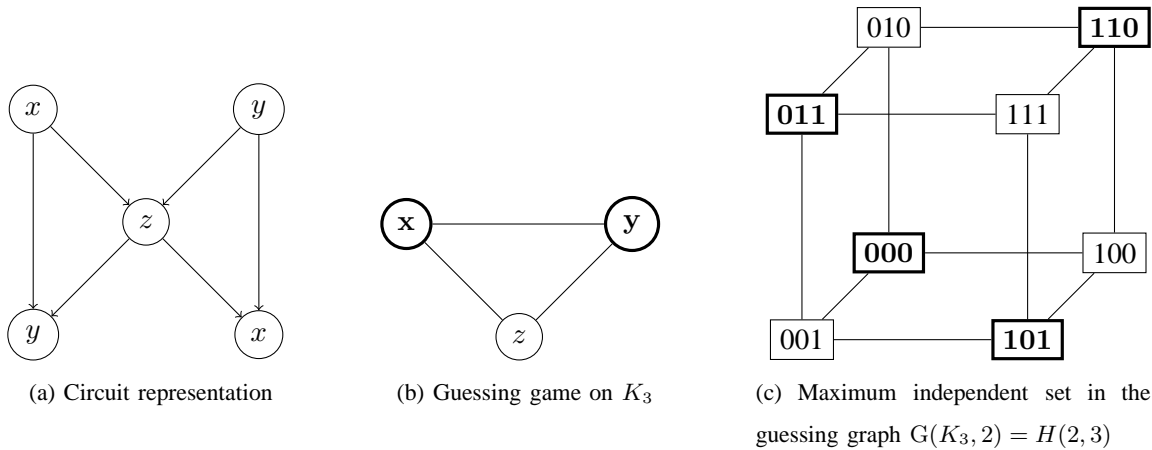

(a) Circuit representation  (b) Guessing game on $K_3$  (c) Maximum independent set in the guessing graph $\mathrm{G}(K_3, 2) = H(2, 3)$

Fig. 2. The butterfly network as a maximum independent set problem.

we shall represent the configurations in rectangular vertices and shall highlight a maximum independent set in bold with thick contours.

*B. Results based on the guessing graph*

We now investigate the properties of the guessing graph and thus derive bounds on the guessing number and on the information defect of digraphs. We first show in Proposition 2 below that the general guessing number and the general information defect of a digraph are equivalent. From a guessing game perspective, this shows that the minimum amount of information required to guess everything correctly ($b(D)$) is exactly equal to the amount of information that is not inferred by the players ($n - g(D)$).

*Proposition 2:* For any digraph $D$, we have $b(D) + g(D) = n$.

*Proof:* The bounds on the chromatic number and the independence number of a vertex transitive graph in (2) yield $b(D, s) + g(D, s) \geq n$ and for $s \geq 3$

$$\begin{aligned} b(D, s) &\leq n - g(D, s) + \log_s(1 + g(D, s) \log s) \\ &\leq n - g(D, s) + \log_s n + \log_s \log s, \end{aligned}$$

which asymptotically yields $b(D) = n - g(D)$.  ∎

Remark that the equality $b(D, s) + g(D, s) = n$ may not hold for all digraphs and every $s$ (e.g., the undirected pentagon over alphabets with $s$ non-square [11]). However, it does hold for every $s$ for the digraphs considered in Examples 1 and 2.

The following proposition gives a lower bound on the guessing number based on the degree of the guessing graph, which shall be refined for large alphabets in Proposition 5.

*Proposition 3:* For any non-acyclic digraph $D$ with minimum in-degree $\delta$ and any $s$,

$$g(D,s) \geq n + \log_s \frac{3}{2} + \log_s \left\{ 1 - \sqrt{1 - \frac{4}{3(d(G(D,s)) + 1)}} \right\} \geq \delta - \log_s n.$$

*Proof:* Since the guessing graph is vertex-transitive, its connectivity is at least $\frac{2(d+1)}{3}$ by [32]. By applying the first inequality in (1), we easily obtain the first lower bound above. Call this term $L$; the second inequality in (1) yields $L \geq n - \log_s(d(G(D,s)) + 1)$. We have $d(G(D,s)) = |\bigcup_i E_i \cap \{0\}|$, where $|E_i \cap \{0\}| = (s-1)s^{n-d_i-1}$ as seen in the proof of Proposition 1, and hence $d(G(D,s)) \leq ns^{n-\delta} - 1$. The second lower bound then follows. ■

If $H$ is a spanning subgraph of $D$, then it is easy to verify that $G(H,s) \supseteq G(D,s)$, and hence $g(H,s) \leq g(D,s)$. Intuitively, $H$ is obtained from $D$ by removing edges, hence less information can circulate. On the other hand, the guessing graph of any induced subgraph can be viewed as a subgraph of the guessing graph of $D$. For any induced subgraph $H$ of $D$ and any $e \in [s]^{n-|H|}$, we denote the subgraph of $G(D,s)$ induced by all configurations satisfying $x_{V-H} = e$ as $G(D,s)_H + e$.

*Lemma 1:* For any induced subgraph $H$ of $D$ and any $e \in [s]^{n-|H|}$, we have $G(D,s)_H + e \cong G(H,s)$.

*Proof:* Two configurations $x, y$ are adjacent in $G(D,s)_H + e$ if and only if there exists $v_i \in H$ such that $x_i \neq y_i$, $x_{N_-(v_i)} = y_{N_-(v_i)}$. Since $x_{V-H} = y_{V-H} = e$, this is equivalent to $x_i \neq y_i$, $x_{N_-(v_i) \cap H} = y_{N_-(v_i) \cap H}$, and hence $x_H$ and $y_H$ are adjacent in $G(H,s)$. ■

*Corollary 1:* We have $\log_s \omega(G(D,s)) \geq mas(D)$, where $mas(D)$ denotes the maximum size of an acyclic induced subgraph of $D$.

*Proof:* Let $H$ be a maximum induced acyclic subgraph of $D$, then $G(D,s)_H + e \cong G(H,s)$, which by Example 1 is a clique on $s^{|H|}$ vertices. ■

The proof of Corollary 1 actually indicates that the family $\{G(D,s)_H + e\}$ for all $e \in [s]^{n-mas(D)}$ forms a partition of the vertex set of $G(D,s)$ into cliques of size $s^{mas(D)}$.

Proposition 4 below combines the results derived above with the graph-theoretic results reviewed in Section II-A.

*Proposition 4:* For any non-acyclic digraph $D$ and any $s \geq 2$,

$$\frac{n}{\Delta + 1} \leq mas(D) \leq \log_s \omega(G(D,s))$$
$$\leq \log_s \chi^*(G(D,s)) = n - \log_s \alpha(G(D,s)) = n - g(D,s)$$
$$\leq \log_s \chi(G(D,s)) = b(D,s)$$
$$\leq \log_s d(G(D,s)) \leq n - \delta + \log_s n.$$

A code with Hamming distance $d$ can be viewed as an independent set of the graph where two words are adjacent if and only if they differ by at most $d - 1$ coordinates. Therefore, finding a maximum code with a prescribed minimum distance can be viewed as finding the maximum independent set of this graph. On the other hand, as seen in Proposition 1, whether two configurations are adjacent in the guessing graph is completely determined by the coordinates in which they differ. Therefore, determining the guessing number of a digraph is a similar problem to that of finding error-correcting codes with maximum cardinality. In particular, Example 1 indicates that the guessing number of the clique $K_n$ (the directed cycle $C_n$, respectively) is equivalent to the maximum cardinality of a code of length $n$ with minimum distance 2 (minimum distance $n$, respectively). Proposition 5 generalizes this property by viewing a set of fixed configurations as a code, and by bounding its minimum distance.

*Proposition 5:* If $D$ is a digraph with minimum in-degree $\delta$ and girth $\gamma$, then

$$\log_s A_s(n, n - \delta + 1) \leq g(D, s) \leq \log_s A_s(n, \gamma).$$

In particular, $g(D, s) \geq \delta$ for $s$ the power of a prime and either $s \geq n - 1$ or $s = 2^m$, $n = 2^m + 2$, and $\delta \in \{4, 2^m\}$ for some $m$.

*Proof:* First, for any two configurations $x, y \in [s]^n$ adjacent in the guessing graph of $D$, we have $(x - y)_{N_-(v_i)} = 0$ for some $i$, and hence $d_{\mathrm{H}}(x, y) \leq n - d_i \leq n - \delta$. Therefore, in any code with minimum distance $n - \delta + 1$, the codewords are not adjacent in the guessing graph, and hence they form a set of fixed configurations.

Conversely, let $x, y$ be two distinct configurations which are not adjacent in the guessing graph, and denote $I = \{v_i : x_i \neq y_i\}$ so that $x, y \in \mathrm{G}(D, s)_I + x_{V-I}$. Suppose $I$ is acyclic, then $\mathrm{G}(I, s)$ is a clique by Example 1, and by Lemma 1, $\mathrm{G}(D, s)_I + x_{V-I}$ is also a clique, and hence $x$ and $y$ are adjacent in $\mathrm{G}(D, s)$. This is a contradiction, thus $I$ contains a cycle and its cardinality is no less than the girth of $D$. Therefore, the set of fixed configurations of any protocol is a code with minimum distance at least $\gamma$.

Since any code with minimum Hamming distance $n - \delta + 1$ forms a set of fixed configurations, using an MDS code yields the lower bound $g(D, s) \geq \delta$ for the mentioned parameter values. ∎

Proposition 5 implies that for large enough alphabets, the smallest amount $\delta$ of information received by any vertex can circulate through the network.

*C. Combining two graphs*

We now investigate how to combine two digraphs $H_1$ and $H_2$ with disjoint vertex sets. We consider three different types of digraph union, each leading to a different graph product of their guessing graphs.

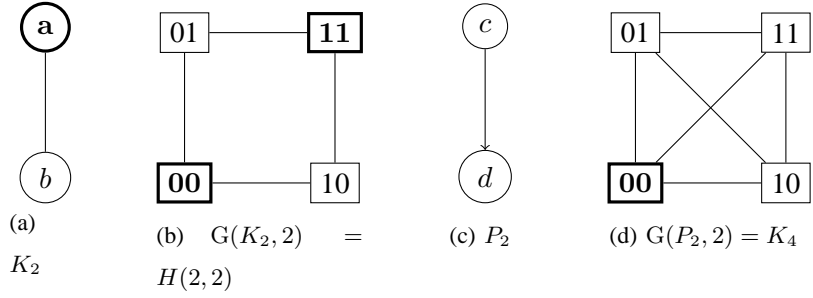

Fig. 3. The digraphs $K_2$ and $P_2$ and their guessing graphs.

We shall illustrate these unions by the following example: $H_1 = K_2$ and $H_2 = P_2$ illustrated in Figure 3.

First, the *disjoint union* of $H_1$ and $H_2$, denoted as $H_1 \cup H_2$, has $V(H_1) \cup V(H_2)$ as vertex set and $E(H_1) \cup E(H_2)$ as edge set. Its adjacency matrix is hence given by

$$\mathbf{A}_{H_1 \cup H_2} = \left( \begin{array}{c|c} \mathbf{A}_{H_1} & \mathbf{0} \\ \hline \mathbf{0} & \mathbf{A}_{H_2} \end{array} \right).$$

In other words, the digraphs are simply placed next to each other, without adding any edges. For any $D$ with vertex set $V(D) = V(H_1) \cup V(H_2)$, we have $D \supseteq H_1 \cup H_2$ and hence the guessing number of the disjoint union of $H_1$ and $H_2$ is a lower bound for the guessing number of $D$. In [12, Lemma 3.2], it is shown that the (linear) guessing number of the disjoint union of two digraphs is equal to the sum of their (linear) guessing numbers. We give an alternate proof below for the nonlinear case by considering the guessing graphs.

*Proposition 6:* For all digraphs $H_1$, $H_2$ with disjoint vertex sets and any $s \geq 2$,

$$\mathrm{G}(H_1 \cup H_2, s) \cong \mathrm{G}(H_1, s) \oplus \mathrm{G}(H_2, s), \tag{5}$$

where $\oplus$ denotes the co-normal product, and hence $g(H_1 \cup H_2, s) = g(H_1, s) + g(H_2, s)$.

*Proof:* Let $x$ and $y$ be two configurations on $H_1 \cup H_2$, and denote $x_{H_1} = x^1$, $y_{H_1} = y^1$ (and similarly for $H_2$). They are adjacent in $\mathrm{G}(H_1 \cup H_2, s)$ if and only if there exists $v_i$ in $H_1$ or in $H_2$ such that $x_i \neq y_i$ and $x_{N_-(v_i)} = y_{N_-(v_i)}$. Since the neighborhood of $v_i$ entirely lies in $H_1$ if $v_i \in H_1$ (and similarly for $H_2$), this is equivalent to $x_i^1 \neq y_i^1$, $x_{N_-(v_i)}^1 = y_{N_-(v_i)}^1$ or $x_i^2 \neq y_i^2$, $x_{N_-(v_i)}^2 = y_{N_-(v_i)}^2$. Therefore, this is equivalent to $x^1 \sim y^1$ in $\mathrm{G}(H_1, s)$ or $x^2 \sim y^2$ in $\mathrm{G}(H_2, s)$, which yields (5). Finally, (3) gives the guessing number of the disjoint union. ∎

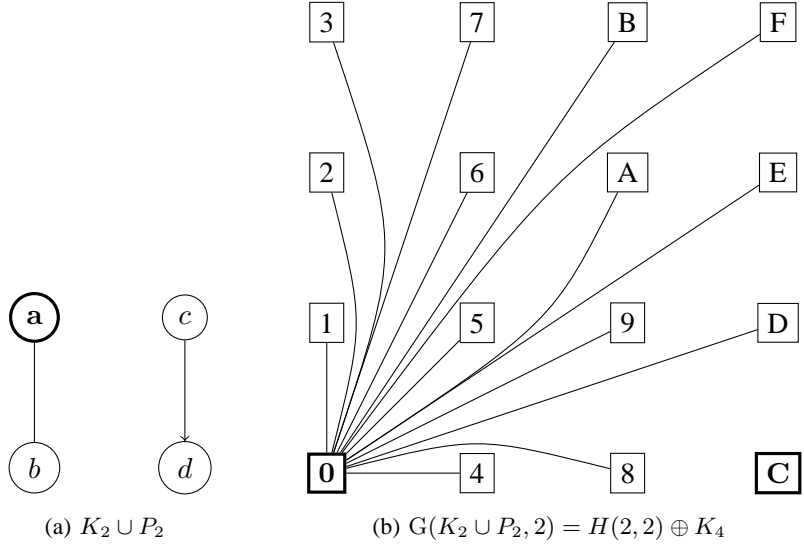

(a) $K_2 \cup P_2$

(b) $G(K_2 \cup P_2, 2) = H(2,2) \oplus K_4$

Fig. 4. The disjoint union of $K_2$ and $P_2$ and its guessing graph.

*Example 3:* The guessing graph of the disjoint union of $K_2$ and $P_2$ is illustrated in Figure 4 below (we represent the configurations in hexadecimal form). Because it is a very dense graph, we only show which configurations are adjacent to the all-zero configuration. It is clear that $\alpha(G(K_2 \cup P_2), 2) = 2$ and hence $g(K_2 \cup P_2, 2) = 1$.

As a corollary of Proposition 6, we now give lower bounds on the guessing number of a digraph by considering the sum of guessing numbers of its induced subgraphs. We refer to a *clique partition* as a partition of the vertex set of a digraph into $r$ subsets such that the graph induced by each subset forms a clique. The *clique partition number* of a digraph $D$, denoted as $c(D)$, is the minimum number of subsets in any clique partition of $D$. Then it is easily shown that $g_{\text{linear}}(D, s) \geq n - c(D)$, which actually refines the lower bound in [12, Theorem 3.3] for graphs with bidirectional edges.

We strengthen the result on the guessing number of the disjoint union below by considering the *unidirectional union* of $H_1$ and $H_2$, denoted as $H_1 \vec{\cup} H_2$, and defined to be $(V(D), E(D))$ with $V(D) = V(H_1) \cup V(H_2)$ and $E(D) = E(H_1) \cup E(H_2) \cup \{(i,j) : i \in V(H_1), j \in V(H_2)\}$. Its adjacency matrix is given by

$$\mathbf{A}_{H_1 \vec{\cup} H_2} = \left( \begin{array}{c|c} \mathbf{A}_{H_1} & \mathbf{1} \\ \hline \mathbf{0} & \mathbf{A}_{H_2} \end{array} \right).$$

In other words, we make all the possible connections, but only from $H_1$ to $H_2$.

*Proposition 7:* For all $H_1$, $H_2$ with disjoint vertex sets and any $s \geq 2$,

$$\mathrm{G}(H_1 \vec{\cup} H_2, s) \cong \mathrm{G}(H_1, s) \cdot \mathrm{G}(H_2, s),$$

where $\cdot$ is the lexicographic product and hence $g(H_1 \vec{\cup} H_2, s) = g(H_1, s) + g(H_2, s)$. Also, we have $g_{\mathrm{linear}}(H_1 \vec{\cup} H_2, s) = g_{\mathrm{linear}}(H_1, s) + g_{\mathrm{linear}}(H_2, s)$.

*Proof:* The proof for the guessing number is similar to that of Proposition 6, and is hence omitted. We hence prove the result for the linear guessing number. For any $\mathbf{A} \leq \mathbf{A}_{H_1 \vec{\cup} H_2}$, we have

$$\mathbf{I}_n + \mathbf{A} = \left( \begin{array}{c|c} \mathbf{I}_{n_1} + \mathbf{A}_1 & \mathbf{A}_3 \\ \hline \mathbf{0} & \mathbf{I}_{n_2} + \mathbf{A}_2 \end{array} \right),$$

where $\mathbf{A}_1 \leq \mathbf{A}_{H_1}$ and $\mathbf{A}_2 \leq \mathbf{A}_{H_2}$. Therefore,

$$\begin{aligned} \mathrm{rk}(\mathbf{I}_n + \mathbf{A}) &\geq \mathrm{rk}(\mathbf{I}_{n_1} + \mathbf{A}_1) + \mathrm{rk}(\mathbf{I}_{n_2} + \mathbf{A}_2) \quad (6) \\ &\geq \min_{\mathbf{A}_1 \leq \mathbf{A}_{H_1}} \mathrm{rk}(\mathbf{I}_{n_1} + \mathbf{A}_1) + \min_{\mathbf{A}_2 \leq \mathbf{A}_{H_2}} \mathrm{rk}(\mathbf{I}_{n_2} + \mathbf{A}_2), \end{aligned}$$

and hence $g_{\mathrm{linear}}(H_1 \vec{\cup} H_2, s) \leq g_{\mathrm{linear}}(H_1, s) + g_{\mathrm{linear}}(H_2, s)$ by (4). Furthermore, if $\mathbf{A}_3 = \mathbf{0}$, we have equality in (6) and hence we can easily prove the reverse inequality. ∎

*Example 4:* The guessing graph of the unidirectional union of $K_2$ and $P_2$ is illustrated in Figure 5 below. Because it is a very dense graph, we only show which configurations are adjacent to the all-zero configuration. Although it is distinct to the guessing graph of the disjoint union, they both have the same independence number.

Proposition 7 indicates that the edges from $H_1$ to $H_2$ do not increase the guessing number and can hence be omitted. Intuitively, the edges only going in one direction, they do not create any more cycles, and hence no more information can circulate through the whole digraph. If we apply this simplification recursively, we obtain that the guessing number of a digraph is completely determined by the guessing numbers of its strong components.

*Corollary 2:* For any digraph $D$ with strong components $C_i$ for $1 \leq i \leq r$, we have $g(D, s) = \sum_{i=1}^{r} g(C_i, s)$ and $g_{\mathrm{linear}}(D, s) = \sum_{i=1}^{r} g_{\mathrm{linear}}(C_i, s)$. Therefore, $g(D, s) \leq n - r$.

*Proof:* The proof goes by induction on the number $r$ of strong components. The case where $r = 1$ is straightforward. Let us assume the result is true for all digraphs with at most $r - 1$ components and consider $D$ with $r$ components. It is well-known that if each component is contracted to a single vertex, the resulting digraph, referred to as the condensation of $D$, is acyclic. In this condensation, there exists a vertex with in-degree 0 (without loss, corresponding to the component $C_1$) such that $D = C_1 \vec{\cup} H$, where

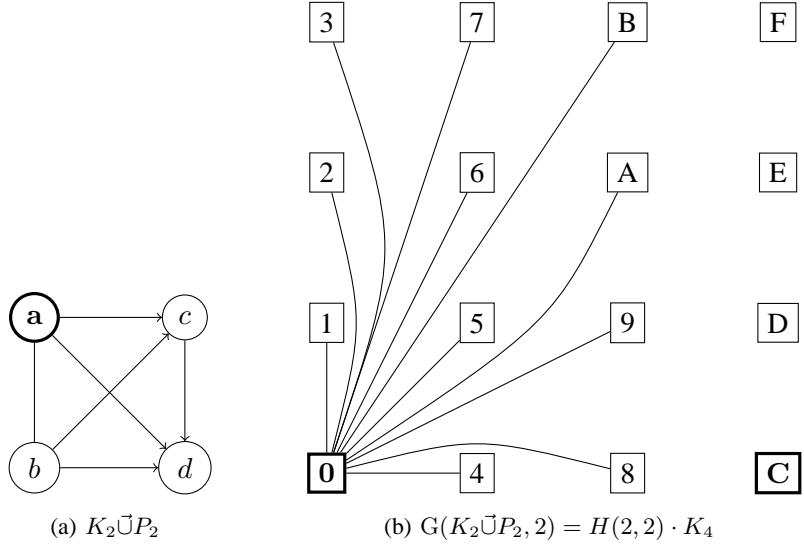

(a) $K_2 \vec{\cup} P_2$    (b) $G(K_2 \vec{\cup} P_2, 2) = H(2, 2) \cdot K_4$

Fig. 5. The unidirectional union of $K_2$ and $P_2$ and its guessing graph.

$H$ is the subgraph induced by $V(D) - V(C_1)$. We then have $g(D, s) = g(C_1, s) + g(H, s)$; however, since $H$ has $r - 1$ components $C_2, \ldots, C_r$, we obtain $g(D, s) = g(C_1, s) + g(C_2, s) + \ldots + g(C_r, s)$. The proof is similar for the linear case. Finally, since $g(C_i, s) \leq |C_i| - 1$ for all $i$, we have $g(D, s) \leq n - r$. ∎

Finally, the *bidirectional union* of two digraphs, denoted as $H_1 \bar{\cup} H_2$, is obtained by connecting all vertices of $H_1$ to those of $H_2$, and vice versa. We have $E(H_1 \bar{\cup} H_2) = E(H_1) \cup E(H_2) \cup \{(i_1, i_2), (i_2, i_1) : i_1 \in V(H_1), i_2 \in V(H_2)\}$. Its adjacency matrix is given by

$$\mathbf{A}_{H_1 \bar{\cup} H_2} = \left( \begin{array}{c|c} \mathbf{A}_{H_1} & \mathbf{1} \\ \hline \mathbf{1} & \mathbf{A}_{H_2} \end{array} \right).$$

Clearly, for any digraph $D$ and any two induced subgraphs $H_1$ and $H_2$ of $D$ with disjoint vertex sets, we have $D \subseteq H_1 \bar{\cup} H_2$; therefore, the guessing number of the bidirectional union is an upper bound on the guessing number of any union of $H_1$ and $H_2$.

*Proposition 8:* For any $H_1$, $H_2$ with disjoint vertex sets and any $s \geq 2$,

$$G(H_1 \bar{\cup} H_2, s) \cong G(H_1, s) \square G(H_2, s),$$

where $\square$ denotes the cartesian product. Therefore,

$$b(H_1 \bar{\cup} H_2, s) = \max\{b(H_1, s), b(H_2, s)\}, \tag{7}$$
$$g(H_1 \bar{\cup} H_2, s) \leq \min\{g(H_1, s) + n_2, g(H_2, s) + n_1\}.$$

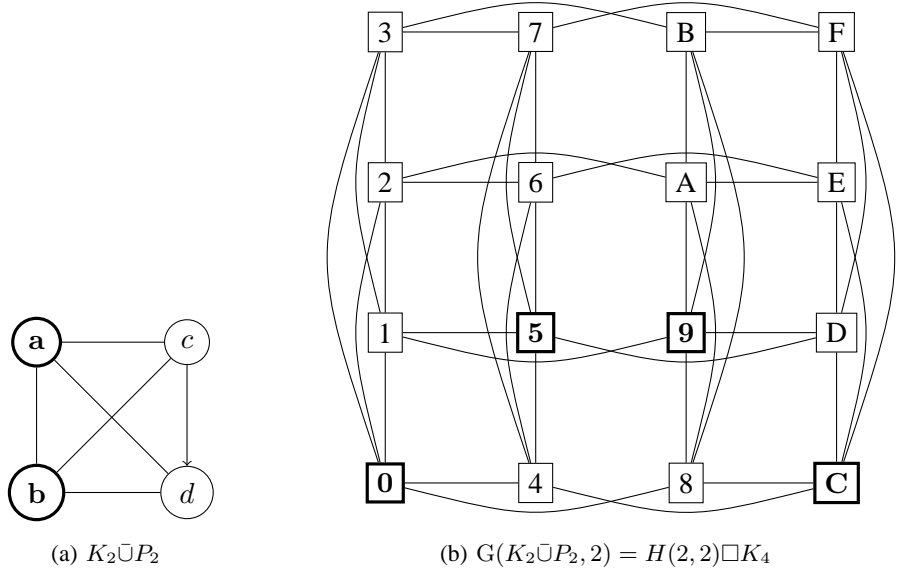

Fig. 6. The bidirectional union of $K_2$ and $P_2$ and its guessing graph.

In the linear case, we have $g_{\text{linear}}(H_1 \dot{\cup} H_2, s) = \min\{g_{\text{linear}}(H_1, s) + n_2, g_{\text{linear}}(H_2, s) + n_1\}$.

*Proof:* The proof for the general case is similar to that of Proposition 6 and hence omitted. We now prove the linear case. Let $\mathbf{A} \leq \mathbf{A}_{H_1 \dot{\cup} H_2}$ such that $\text{rk}(\mathbf{I}_n + \mathbf{A}) = n - g_{\text{linear}}(H_1 \dot{\cup} H_2, s)$. Since

$$\mathbf{I}_n + \mathbf{A} = \left(\begin{array}{c|c} \mathbf{I}_{n_1} + \mathbf{A}_1 & \mathbf{A_3} \\ \hline \mathbf{A}_4 & \mathbf{I}_{n_2} + \mathbf{A}_2 \end{array}\right)$$

for some $\mathbf{A}_1 \leq \mathbf{A}_{H_1}$ and $\mathbf{A}_2 \leq \mathbf{A}_{H_2}$, we have $\text{rk}(\mathbf{I}_n + \mathbf{A}) \geq \max\{\text{rk}(\mathbf{I}_n + \mathbf{A}_1), \text{rk}(\mathbf{I}_n + \mathbf{A}_2)\} \geq \max\{n_1 - g_{\text{linear}}(H_1, s), n_2 - g_{\text{linear}}(H_2, s)\}$.

Conversely, without loss suppose $l = n_1 - g_{\text{linear}}(H_1, s) \geq n_2 - g_{\text{linear}}(H_2, s)$ and let $\mathbf{A}_1$ and $\mathbf{A}_2$ satisfy $\text{rk}(\mathbf{A}_i) = n_i - g_{\text{linear}}(H_i)$ for $i = 1, 2$. We can express $\mathbf{A}_i$ as $\mathbf{A}_i = \mathbf{B}_i^T \mathbf{C}_i$, where $\mathbf{B}_i, \mathbf{C}_i \in \text{GF}(s)^{l \times n_i}$. Then the matrix $\mathbf{A} = (\mathbf{B}_1, \mathbf{B}_2)^T (\mathbf{C}_1, \mathbf{C}_2)$ has rank $l$. ∎

*Example 5:* The guessing graph of the bidirectional union of $K_2$ and $P_2$ is depicted in Figure 6 below. In this case, we have $g(K_2 \dot{\cup} P_2, 2) = g(P_2, 2) + 2$ because the optimal protocols are linear.

*Example 6:* Consider the following network coding instance, where $n$ sources want to transmit a message each via a common bottleneck of $m \leq n$ nodes (depicted in Figure 7 for $n = 3$, $m = 2$). The network coding is solvable if and only if the complete bipartite graph $K_{m,n}$ has guessing number $n$. Since this digraph can be viewed as the bidirectional union of the empty graphs on $n$ and $m$ vertices, its guessing number is upper bounded by $m$ by Proposition 8. Conversely, since it contains $m$ disjoint

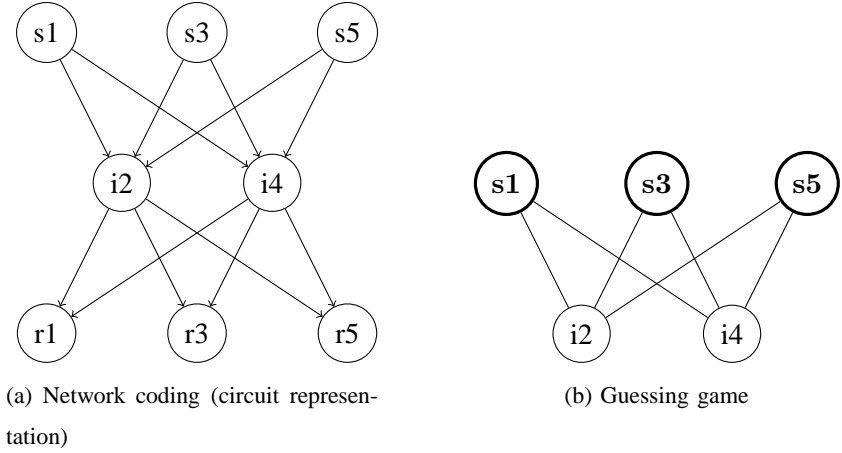

(a) Network coding (circuit representation)

(b) Guessing game

Fig. 7. The bottleneck with $n = 3$, $m = 2$.

cliques $K_2$, its guessing number is lower bounded by $m$. Therefore, the network coding instance in Figure 7 is solvable if and only if $m = n$, i.e., there is no bottleneck and routing is sufficient.

*D. Combining alphabets*

A network coding instance solvable over $[s]$ is clearly solvable over $[s^k]$ for any $k \geq 2$. However, it is shown in [33] that certain network coding instances can be solvable over an alphabet but not over some larger alphabet. In this section, we discover interesting properties of the guessing graphs of the same digraph over different alphabets, which yield bounds on and relations amongst the guessing numbers of a digraph over different alphabets. First, a set of fixed configurations of a protocol on $D$ over $[s]$ can also be viewed as fixed configurations of a protocol over the alphabet $[t]$, for any $t \geq s$ which yields

$$g(D,t) \geq g(D,s) \log_t s. \tag{8}$$

We refine this bound below by showing that the guessing graph on the cartesian product of two alphabets is closely related to the guessing graphs on the two initial alphabets.

*Proposition 9:* For any digraph $D$ and any $s, t \geq 2$ we have

$$\mathrm{G}(D,s) \square \mathrm{G}(D,t) \subseteq \mathrm{G}(D,st) \subseteq \mathrm{G}(D,s) \oplus \mathrm{G}(D,t), \tag{9}$$

and hence

$$\frac{g(D,s) \log s + g(D,t) \log t}{\log s + \log t} \leq g(D,st) \leq \min \left\{ \frac{g(D,s) \log s + n \log t}{\log s + \log t}, \frac{g(D,t) \log t + n \log s}{\log s + \log t} \right\}.$$

*Proof:* Since the sets $[st]$ and $[s] \times [t]$ are isomorphic, we consider two configurations $(x^s, x^t), (y^s, y^t) \in ([s] \times [t])^n$. Suppose they are adjacent in $\mathrm{G}(D, st)$; therefore there exists $i$ such that $(x_i^s, x_i^t) \neq (y_i^s, y_i^t)$ and $(x_{N_-(v_i)}^s, x_{N_-(v_i)}^t) = (y_{N_-(v_i)}^s, y_{N_-(v_i)}^t)$. This is equivalent to $x_{N_-(v_i)}^s = y_{N_-(v_i)}^s$ and $x_{N_-(v_i)}^t = y_{N_-(v_i)}^t$ and ($x_i^s \neq y_i^s$ or $x_i^t \neq y_i^t$). It is easy to check that they are adjacent in $\mathrm{G}(D, s) \oplus \mathrm{G}(D, t)$. Moreover, we can similarly prove the other inclusion. ∎

As a corollary, we obtain that the guessing number over any alphabet can serve as a lower bound for the guessing numbers over larger alphabets.

*Corollary 3:* For any $t \geq s$ with $m = \lfloor \log_s t \rfloor$, we have

$$g(D,s) \frac{m}{\log_s t} \leq g(D,t) \leq \frac{g(D,s) + mn}{\log_s t}.$$

*Proof:* By applying Proposition 9 recursively, we obtain $g(D, s^{m+1}) \leq \frac{g(D,s) + mn}{m+1}$, and the upper bound follows from (8). Also, applying (9) recursively yields $g(D, s^l) \geq g(D, s)$ for all $l \geq 1$, which combined with (8) yields the lower bound. ∎

The result in (9) can be interpreted using digraph unions. For any digraph $D$ and any $k \geq 1$, we denote the digraph $k \oplus D$, whose vertex set is given by $V(k \oplus D) = \{\mathbf{v} = (v, i) : v \in V(D), i \in [k]\}$ and whose edge set is $E(k \oplus D) = \{(\mathbf{u}, \mathbf{v}) : (u, v) \in E(D)\}$. In other words, we take $k$ copies of $D$ and make connections between the copies corresponding to the edges in $D$. Therefore, the in-neighborhood of a vertex $(v, i)$ in $k \oplus D$ consists of the $k$ copies of the in-neighborhood of $v$. In terms of network coding, the digraph $k \oplus D$ can be viewed as expanding the instance according to the $k$ symbols in $[s]$ of an element of $[s^k]$.

*Proposition 10:* For any $D$, $k$, and $s$, we have $\mathrm{G}(k \oplus D, s) = \mathrm{G}(D, s^k)$ and hence $g(k \oplus D, s) = kg(D, s^k)$.

The proof is similar to that of Proposition 6 and is hence omitted. Note that for $k = 2$ and $D_1 \cong D_2 \cong D$, we have $D_1 \cup D_2 \subseteq 2 \oplus D \subseteq D_1 \bar{\cup} D_2$; hence (9) can be viewed as an extension of Proposition 10 to mixed alphabets. Proposition 10 means that playing the guessing game over extension fields is equivalent to playing the guessing game over the base field, but on several copies of the digraph.

The result in Proposition 10 also implies that $2 \oplus D$ is the union of two copies of $D$ which, like the unidirectional union of Proposition 7, does not improve on the general guessing number of the disjoint union. As seen before, the unidirectional union did not add any cycles to the digraph, hence the information could not circulate between the two copies of the digraph. On the other hand, the union $2 \oplus D$ does create new cycles, yet the information received by any vertex is redundant as the in-neighborhood of any vertex in $2 \oplus D$ is simply two copies of its in-neighborhood in $D$. For instance, the digraph $2 \oplus C_3$ illustrated in Figure 8 has guessing number 2 over any alphabet.

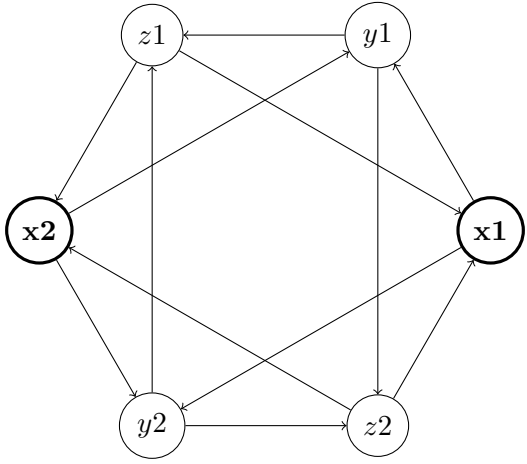

Fig. 8.  The digraph $2 \oplus C_3$ with guessing number 2.

## IV. A CONSTRUCTION OF DIGRAPHS BASED ON CYCLIC CODES

In this section, for the sake of simplicity we only consider the *binary* guessing number (i.e., $s = 2$). However, the concepts introduced below can be easily extended to any field.

### A. Digraphs generated by cyclic codes

We first define a simple linear protocol which takes advantage of all the information incoming at every node.

*Definition 2:* The *parity-check protocol* $\mathcal{H}$ has the functions $\mathcal{H}(x)_v$ defined for any $v \in V$ as $f_v(x_{N_-(v)}) = \mathbf{1} \cdot x_{N_-(v)}$, or equivalently $f_v(x_{N_-(v)}) = \sum_{v_j \in N_-(v)} x_{v_j}$.

By definition, the parity-check protocol is linear, hence its fixed configurations form a linear binary code. It is easily shown that it has an extended parity-check matrix given by $\mathbf{H}' = \mathbf{I}_n + \mathbf{A}_D^T$. Clearly, the rows of $\mathbf{H}'$ may be linearly dependent, as seen in Example 7 below. Therefore, our aim is to use extended parity check matrices with low rank.

*Example 7:* Let $C_3$ be the directed cycle on three edges with adjacency matrix
$$\mathbf{A}_D = \begin{pmatrix} 0 & 1 & 0 \\ 0 & 0 & 1 \\ 1 & 0 & 0 \end{pmatrix}.$$

The resulting matrix $\mathbf{H}'$ is given by

$$\mathbf{H}' = \begin{pmatrix} 1 & 0 & 1 \\ 1 & 1 & 0 \\ 0 & 1 & 1 \end{pmatrix},$$

which has rank 2. Therefore, the fixed configurations of the parity-check protocol form a $(3, 1)$ binary code (the repetition code) whose generator matrix is given by

$$\mathbf{G} = \begin{pmatrix} 1 & 1 & 1 \end{pmatrix}.$$

Any linear protocol on a digraph $D$ can be viewed as the parity-check protocol on a subgraph of $D$. Therefore, the linear guessing number of $D$ is given by the logarithm of the maximum number of fixed configurations of the parity-check protocol over all subgraphs of $D$. In other words, we do not lose any generality by considering the parity-check protocol only instead of any linear protocol. The maximum linear guessing number over all digraphs with no bidirectional edges is hence given by the logarithm of the maximum number of fixed configurations of the parity-check protocol of all digraphs with no bidirectional edges.

We now reverse the problem, and construct digraphs based on linear codes. Clearly, any collection of vectors $c_0, c_1, \ldots, c_{n-1} \in \mathrm{GF}(2)^n$ where the $i$-th coordinate of $c_i$ is equal to 1 would produce a matrix of the type $\mathbf{I} + \mathbf{A}_D$ for some digraph $D$, and the code would simply be the dual of the span of these vectors. Since the properties of the obtained digraph are not easy to determine in general, we focus on the class of cyclic codes.

*Definition 3:* Let $C$ be an $(n, k)$ binary cyclic code generated by the polynomial $g(x)$. Then the digraph generated by $C$ has adjacency matrix $\mathbf{I}_n + \mathbf{H}'^T$, where the rows of $\mathbf{H}'$ are the $n$ cyclic shifts of $g(x)$. Equivalently, denoting $g(x) = \sum_{i=0}^{n-1} g_i x^i$, there is an edge from $v_{a+i \mod n}$ to $v_a$ if and only if $g_i = 1$ for all $a$ and $i$.

*Example 8:* Three trivial polynomials generate the following digraphs.

- The polynomial $g(x) = 1$ generates the empty graph;
- $g(x) = x + 1$ generates the directed cycle $C_n$ (in particular, $C_3$ given in Example 7 is generated by the $(3, 2)$ single parity-check code);
- $g(x) = \frac{x^n + 1}{x + 1} = x^{n-1} + x^{n-2} + \ldots + 1$ generates the clique $K_n$.

The generation of the clique can be generalized when $n = st$ is a composite number. Then we have $x^{st} + 1 = (x^s + 1)(x^{(t-1)s} + x^{(t-2)s} + \ldots + x^s + 1)$, hence the rightmost polynomial generates an $(st, s(t-1))$ cyclic code, which generates the disjoint union of $s$ cliques of size $t$ each. According to our previous results, this digraph has in-degree and out-degree equal to $t-1$, while its linear guessing number is $s(t-1)$. This digraph is not connected; however, by adding a cycle $C_n$ that connects all the vertices, we make the digraph strong, while increasing the in-degree by 1. We thus obtain a class of strong regular digraphs on $n$ vertices and in-degree $d$ satisfying $g_{\text{linear}}(D, s) \geq n - \frac{n}{d}$ for all values of $d$.

The properties of digraphs generated by cyclic codes are listed in Theorem 2 below.

*Theorem 2:* The digraph $D$ on $n$ vertices generated by $C$ with generator polynomial $g(x) = \sum_{i=0}^{n-1} g_i x^i$ (hence $g(x)$ divides $x^n + 1$) has the following properties.

1) $D$ is regular with in-degree and out-degree $w(g) - 1$, where $w(g)$ is the number of non-zero coefficients of $g(x)$.
2) $D$ has no bidirectional edges if and only if $g_i g_{n-i} = 0$ for all $1 \leq i \leq \lfloor \frac{n}{2} \rfloor$. In particular, if $\deg(g) < \frac{n}{2}$, then $D$ has no bidirectional edges.
3) $D$ is a tournament if and only if $g_i + g_{n-i} = 1$ for all $1 \leq i \leq n-1$.
4) If $g_i g_j = 1$ for some $i, j \in \{1, 2, \ldots, n\}$ relatively prime, then $D$ is strong.
5) The first $n - \deg(g)$ vertices induce a maximum acyclic subgraph.
6) The binary (linear) guessing number of $D$ satisfies $g_{\text{linear}}(D, 2) = g(D, 2) = \deg(g)$.

*Proof:* The matrix obtained by shifting $g(x)$ $n$ times has the following properties. First, $g(x)$ divides $x^n + 1$ hence $g_0 = 1$ and that matrix has ones all over the diagonal, which ensures that it is the adjacency matrix of some digraph $D$. Second, every row and every column has exactly $w(g)$ ones, which yields Property 1). Properties 2) and 3) are easy to prove.

Third, if $g_i g_j = 1$ for some $i, j$ relatively prime, then we have $ai + bj = 1$ for some $a, b \in \mathbb{Z}$, and hence $a'i + b'j \equiv 1 \mod n$ for $0 \leq a', b' < n$. Therefore, there is a path of length $a' + b'$ from the node $v_e$ to the node $v_{e+1 \mod n}$ for all $0 \leq e \leq n-1$. By iteration, there is a path between $v_e$ and $v_f$ for all $0 \leq e, f \leq n-1$ and $D$ is strong.

Finally, we prove the last two properties simultaneously. It is easy to check that the first $n - \deg(g)$ induce a maximum acyclic subgraph in reverse topological order. The dimension of a cyclic code is equal to $n - \deg(g)$, and hence the dimension of its dual is equal to $\deg(g)$ and $g(D, 2) \geq g_{\text{linear}}(D, 2) \geq \deg(g)$. On the other hand, $g(D, 2) \leq n - mas(D) \leq \deg(g)$ by Proposition 4, implying equalities everywhere. ∎

Properties 5) and 6) naturally imply constructions of solvable network coding instances based on cyclic codes, where the first $n - \deg(g)$ vertices of the digraph generated by $C$ are the intermediate nodes, while the remaining $\deg(g)$ vertices are split into sources and sinks. These instances are solvable over GF(2) using the parity-check protocol, and are hence solvable over any alphabet with cardinality equal to a power of 2.

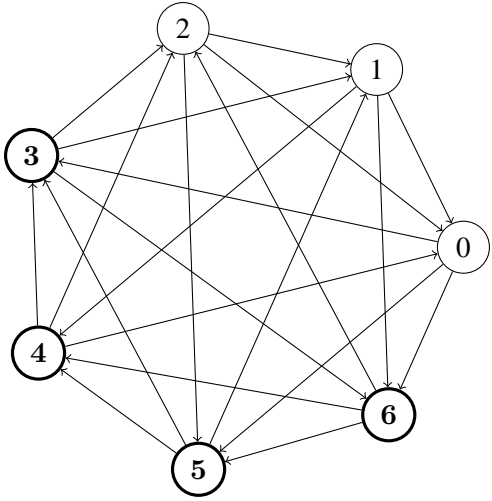

Fig. 9. Digraph $P_7$ on 7 vertices generated by $x^4 + x^2 + x + 1$ with binary linear guessing number 4.

Theorem 2 indicates that a good choice for $g(x)$ has high degree but low weight. We give an example of such a polynomial below.

*Example 9:* Let $n = 7$ and consider the digraph $P_7$ generated by $g(x) = x^4 + x^2 + x + 1$ and illustrated in Figure 9. By Theorem 2, this is a strong and regular tournament, sometimes referred to as a Paley tournament. Its binary linear guessing number is $\deg(g) = 4$, and the fixed configurations form the $(7, 4)$ Hamming code.

This construction illustrates the elegance of the guessing game approach to network coding. Indeed, the source–intermediate node–sink hierarchy in the network coding instance vanishes and all nodes are on the same level, hence yielding more symmetry in the resulting digraph.

More generally, the generator polynomial of the $(2^l - 1, l)$ simplex code generates a digraph on $n_l = 2^l - 1$ vertices, regular with in-degree $d_l = 2^{l-1} - 1$, maximum induced subgraph of size $m_l = l$, and binary linear guessing number $g_l = 2^l - l - 1$. Although these digraphs may have bidirectional edges, the corresponding network coding instances do not. Therefore, we obtain solvable network coding instances where the in-degree is around half the number of vertices, and for which the number of intermediate nodes grows as the logarithm of the number of source-sink pairs.

*B. Digraphs with no bidirectional edges generated by cyclic codes*

So far, we allowed digraphs to have bidirectional edges, which made the search for digraphs with high linear guessing numbers quite easy. We are now interested in digraphs with no bidirectional edges. Based on Theorem 2, this is equivalent to searching for polynomials $g(x)$ dividing $x^n + 1$ such that $g_i g_{n-i} = 0$ for all $1 \leq i \leq \lfloor \frac{n}{2} \rfloor$.

We first give a simple example of such a polynomial. Let $n = 3t$ be a multiple of 3 with $t > 3$, $gcd(t, 3) = 1$, then $x^3 + 1$ and $x^t + 1$ divide $x^n + 1$. In particular, their gcd, given by $(x^t + 1)(x^2 + x + 1) = x^{t+2} + x^{t+1} + x^t + x^2 + x + 1$, is a valid polynomial with degree $t + 2$ and weight 6. Therefore, according to Theorem 2, the digraph generated by this polynomial has in-degree and out-degree 5 and its linear guessing number is $\frac{n}{3} + 2$. Moreover, Theorem 2 ensures that this digraph has no bidirectional edges and is strong.

This example is interesting because it designs a class of digraphs with no bidirectional edges for which we know the linear guessing number is strictly greater than $\frac{n}{3}$. On the other hand, the lower bound in [12, Theorem 3.3] is given by the cycle packing index of the digraph, which can be easily shown to be upper bounded by $\frac{n}{3}$; therefore, that bound is not tight for these digraphs.

If $n = 2p$ is even, then $x^{p-1} + x^{p-2} + \ldots + 1$ is a valid polynomial, which generates a strong unidirectional digraph with in-degree $p - 1$ and whose linear guessing number equal to $p - 1$.

Let $g(x)$ be a factor of $x^{t-1} + x^{t-2} + \ldots + 1 = \frac{x^t}{x+1}$ with degree $d$ and weight $w$. Then for all $l \geq 1$, $x^{2^l t} + 1 = (x^t + 1)^{2^l}$ has $h(x) = (x+1)g^{2^l}(x)$ as factor. The degree of $h(x)$ is clearly $2^l d + 1$, while the weight of $h(x)$ is $2w$, and we have $h_1 = 1$. Therefore, this constructs an infinite class of strong unidirectional digraphs with $2^l t$ vertices, in-degree $2w - 1$, and binary guessing number $2^l d + 1$.

Our approach was restricted to polynomials $g(x)$ which generate a cyclic code, or equivalently, which divide $x^n + 1$. However, any polynomial $h(x)$ where $h_0 = 1$, $h_i h_{n-i} = 0$ for all $i$, and $h_p = 1$ for $p$ relatively prime to $n$ generates a regular strong digraph with no bidirectional edges. The polynomial $h(x)$ belongs to the code generated by the greatest common divisor of $h(x)$ and $x^n + 1$, therefore the guessing number of the digraph generated by $h(x)$ has guessing number lower bounded by $\deg(gcd(h(x), x^n + 1))$.

*Example 10:* Let $n = 14$ and $h(x) = x^{12} + x^{11} + x^{10} + x^9 + x^6 + x + 1$, then

$$gcd(h(x), x^{14} + 1) = x^9 + x^8 + x^6 + x^5 + x^4 + x^3 + 1.$$

In this case, the polynomial has a lower weight than its gcd, and hence sparser digraphs can be generated by considering all polynomials instead of the generator polynomials of cyclic codes only. Nonetheless, considering such general digraphs is not suitable for constructing network coding instances, as the size of a maximum induced subgraph in the digraph generated by $h(x)$ is not easily computable: it is at least $n - \deg(h) = 2$; however, it is actually equal to 3. Note that Theorem 2 does not apply to $h(x)$, as it does not divide $x^n + 1$, and the guessing number is strictly less than the degree of $h(x)$.

## V. ON THE MAXIMUM GUESSING NUMBER OF DIGRAPHS

As seen above, constructing digraphs with high guessing numbers is relatively easy when we allow bidirectional edges. The main purpose of this section is to evaluate the maximum guessing number one obtains when considering strong digraphs with no bidirectional edges. We are particularly interested in the binary linear guessing number of sparse digraphs, which will surprisingly turn out to be sufficient. However, for the sake of completeness, we shall state our results as generally as possible, as some ideas extend to digraphs with bidirectional edges as well.

### A. Upper bounds on the guessing number

We begin this section by deriving upper bounds on the (linear) guessing number of digraphs based on their parameters, such as the minimum or maximum in-degree. We first remark in Lemma 2 that the gap between the guessing number of digraphs and the number of their vertices must grow arbitrarily large. This implies that the probability of success in the guessing game on a digraph with no bidirectional edges tends to zero when the number of players tends to infinity. This also indicates that in any family of solvable network coding instances without any two-hop path between a source and its according sink, the number of intermediate nodes must tend to infinity.

*Lemma 2:* For any digraph $D$ with no bidirectional edges and any $s \geq 2$, we have $g(D, s) \leq n - \log_s((s-1)n + 1)$.

*Proof:* Since $D$ has no bidirectional edges, its girth is at least 3. By Proposition 5, we have $g(D, s) \leq \log_s A_s(n, \gamma) \leq \log_s A_s(n, 3)$. Applying the sphere-packing bound $A_s(n, 3) \leq \frac{s^n}{(s-1)n+1}$, we obtain the desired bound on $g(D, s)$. ∎

Proposition 11 below refines this statement for the linear guessing number of sparse digraphs without bidirectional edges.

*Proposition 11:* For any digraph $D$ on $n$ vertices with no bidirectional edges and with minimum and maximum in-degree $\delta$ and $\Delta$, we have $g_{\text{linear}}(D, s) \leq n - \log_s(n - \delta) - 1$ and $g_{\text{linear}}(D, s) \leq n - \log_s(n - \Delta - e) - 2$, where $e = \max\left\{d : \frac{\binom{n}{\Delta - d + 2}}{\binom{\Delta + 1}{\Delta - d + 2}} \geq n\right\}$.

*Proof:* We first prove the bound based on the minimum in-degree. Let $\mathbf{A} \leq \mathbf{A}_D$ such that $l = \mathrm{rk}(\mathbf{I}_n + \mathbf{A}) = n - g_{\mathrm{linear}}(D, s)$, and denote $\mathbf{B} = \mathbf{I}_n + \mathbf{A}$. Since $D$ has no bidirectional edges, all the rows of $\mathbf{B}$ are distinct. We consider the $s^l$ vectors in the row space of $\mathbf{B}$. Since the fixed configurations of the protocol corresponding to $\mathbf{B}$ form a code with minimum distance at least 2 by the proof of Proposition 5, $s^{l-1}$ vectors have a zero in coordinate $i$ for any $i$. However, let $j$ be a column of $\mathbf{B}$ with at most $\delta + 1$ of ones, i.e. there are at least $n - \delta - 1$ distinct rows of $\mathbf{B}$ with a zero in coordinate $j$, and accounting for the all-zero vector, we obtain $s^{l-1} \geq n - \delta$.

We now prove the bound based on the maximum in-degree. The code with extended parity-check matrix $\mathbf{B}$ has minimum distance at least 3, therefore its dual code (with dimension $l = \mathrm{rk}(\mathbf{B})$) has the following property: for any pair of coordinates $0 \leq i < j \leq n - 1$, $s^{l-2}$ vectors have $(0, 0)$ in these coordinates. Let us give a lower bound on the maximum number, taken over all pairs $\{i, j\}$ of columns, of rows of $\mathbf{B}$ which have $(0, 0)$ in columns $i$ and $j$. First, note that if $\mathbf{C} \leq \mathbf{B}$, then the rows with $(0, 0)$ in $\mathbf{B}$ also have $(0, 0)$ in $\mathbf{C}$. Therefore, without loss, we can assume all the columns of $\mathbf{B}$ have weight $\Delta + 1$. The supports of these columns then form a constant-weight code of length $n$, weight $\Delta + 1$, and cardinality $n$. As seen in Section II-D, its minimum distance $2d$ satisfies $n \leq \binom{n}{\Delta - d + 2} / \binom{\Delta + 1}{\Delta - d + 2}$ and therefore $d \leq e$. Let $i$ and $j$ be two columns of $\mathbf{B}$ at distance $2d$, then the union of their support has cardinality $\Delta + 1 + d$ and there are $n - \Delta - 1 - d$ rows of $\mathbf{B}$ with $(0, 0)$ in coordinates $i$ and $j$. Accounting for the all-zero vector, there are at least $n - \Delta - d$ such vectors, and hence $s^{l-2} \geq n - \Delta - d \geq n - \Delta - e$.
∎

## B. Combining digraphs to increase the guessing number

In Section IV, we showed how to construct digraphs with high guessing numbers for finite parameters. In this section, we investigate how to combine digraphs in order to generate infinite families of digraphs with high guessing numbers.

*Definition 4:* The *strong product* of two digraphs $H_1$ and $H_2$, denoted as $H_1 \boxtimes H_2$ is defined similarly to its counterpart for undirected graphs. Its vertex set is the cartesian product $V(H_1) \times V(H_2)$, and there is an edge from $(u_1, u_2)$ to $(v_1, v_2)$ if and only if either $u_1 = v_1$ and $(u_2, v_2) \in E(H_2)$, or $u_2 = v_2$ and $(u_1, v_1) \in E(H_1)$, or $(u_1, v_1) \in E(H_1)$ and $(u_2, v_2) \in E(H_2)$. Equivalently, the adjacency matrix of the strong product is given by

$$\mathbf{A}_{H_1 \boxtimes H_2} = (\mathbf{I}_{n_1} + \mathbf{A}_{H_1}) \otimes (\mathbf{I}_{n_2} + \mathbf{A}_{H_2}) - \mathbf{I}_{n_1 n_2},$$

where $\otimes$ denotes the Kronecker product of matrices.

The properties of the strong product are listed in Proposition 12 below.

*Proposition 12:* Let $H_1$ and $H_2$ be two digraphs on $n_1$ and $n_2$ vertices, respectively. Then their strong product $H_1 \boxtimes H_2$ has the following properties:

- It has $n = n_1 n_2$ vertices.
- If $H_1$ and $H_2$ are both strong and without any bidirectional edges, then so is $H_1 \boxtimes H_2$.
- If $H_1$ and $H_2$ have regular in-degrees and out-degrees, it is regular with in-degree and out-degree $d(H_1 \boxtimes H_2) = (d(H_1) + 1)(d(H_2) + 1) - 1$.
- Its linear guessing number satisfies $g_{\text{linear}}(H_1 \boxtimes H_2, s) \geq n - (n_1 - g_{\text{linear}}(H_1, s))(n_2 - g_{\text{linear}}(H_2, s))$ for all $s$.

*Proof:* The first three properties are easy to verify. We hence prove the lower bound on the linear guessing number. Let $\mathbf{A}_i \leq \mathbf{A}_{H_i}$ such that $\text{rk}(\mathbf{I}_{n_i} + \mathbf{A}_i) = n_i - g_{\text{linear}}(H_i, s)$ for $i = 1, 2$. Then $(\mathbf{I}_{n_1} + \mathbf{A}_1) \otimes (\mathbf{I}_{n_2} + \mathbf{A}_2) \leq (\mathbf{I}_{n_1} + \mathbf{A}_{H_1}) \otimes (\mathbf{I}_{n_2} + \mathbf{A}_{H_2}) = \mathbf{I}_n + \mathbf{A}_{H_1 \boxtimes H_2}$, which yields $g_{\text{linear}}(H_1 \boxtimes H_2, s) \geq n - \text{rk}\{(\mathbf{I}_{n_1} + \mathbf{A}_1) \otimes (\mathbf{I}_{n_2} + \mathbf{A}_2)\} = n - (n_1 - g_{\text{linear}}(H_1, s))(n_2 - g_{\text{linear}}(H_2, s))$. ∎

*Example 11:* For any $k \geq 1$ and $l \geq 3$, denote the unidirectional cycle $C_l$ raised to the power of $k$ according to the strong product as $C_l^k$ (for instance, $C_3^2$ is illustrated in Figure 10). Then $C_l^k$ is a strongly regular digraph on $n_{l,k} = l^k$ vertices with in-degree and out-degree $d_{l,k} = 2^k - 1$ and linear guessing number $g_{l,k} = l^k - (l-1)^k$. The lower bound on the guessing number follows Proposition 12. The upper bound follows $g(C_l^k, s) \leq n - mas(C_l^k)$ in Proposition 4, where $mas(C_l^k) = (l-1)^k$ since the vertices in $\{0, 1, \ldots, l-2\}^k$ induce an acyclic subgraph.

This yields the following construction of network coding instance. The vertices in $\{0, 1, \ldots, l-2\}^k$ induce an acyclic subgraph, therefore we use them as intermediate nodes. The source and sink nodes come from the split of the other $l^k - (l-1)^k$ vertices of $C_l^k$. Since the linear guessing number is equal to the number of sources, this network coding instance is solvable over any alphabet by linear operations.

The sequences $C_l^k$ for a fixed $l$ have the following property: the ratio between the guessing number over the number of vertices, given by $\frac{g_{\text{linear}}(C_l^k, s)}{n_{l,k}} = 1 - \left(\frac{l-1}{l}\right)^k$ tends to 1 as $k$ tends to infinity. We remark that the convergence could be sped up by considering powers of the digraph $P_7$ depicted in Figure 9, thus obtaining a ratio of $1 - \left(\frac{3}{7}\right)^k$ for alphabets of cardinality equal to a power of 2, but not necessarily for other alphabets.

A consequence of Proposition 4 is that for any family of digraphs with ratio between the guessing number and the number of vertices tending to 1, the maximum in-degree must tend to infinity. On the other hand, the digraphs $C_l^k$ become more and more sparse as $l$ and $k$ increase, as $d_{l,k} + 1 = n_{l,k}^{\log_l 2}$, and hence we can easily construct sequences of strong digraphs with regular in-degree on the order of $n^\epsilon$ for any $\epsilon > 0$. In Theorem 3 below, we strengthen this result by constructing strong digraphs with the ratio of the guessing number over the number of vertices tending to 1 and in-degree tending to infinity as slow as possible.

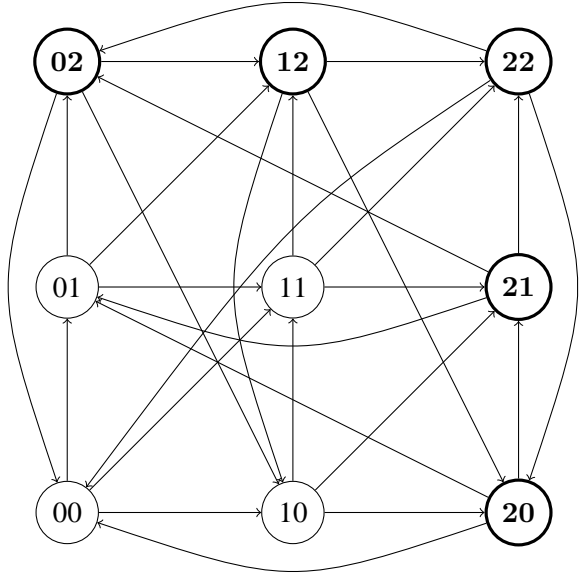

Fig. 10. The digraph $C_3^2$, with linear guessing number 5.

*Theorem 3:* For any $l \geq 3$ and any function $f(n)$ of $n \geq 1$ tending to infinity, there exists an infinite family of strong digraphs $D_k$ on $n_k$ vertices (nondecreasing $n_k$ sequence) with girth $l$ and regular in-degree and out-degree $d_k$ such that $d_k \leq f(n_k)$ for all $k \geq 1$ and $\lim_{k \to \infty} \frac{g_{\text{linear}}(D_k, s)}{n_k} = 1$ for any $s \geq 2$.

*Proof:* For all $k$, let $n_k$ be the smallest multiple of $l^k$ such that $f(n) \geq 2^k$ for all $n \geq n_k$. Then select $m_k = \frac{n_k}{l^k}$ copies of $C_l^k$ and join them by tying a directed cycle around all the vertices. The cycle goes across the different copies as follows. Sort the vertices of $C_l^k$ in lexicographic order, so that $(v_i, v_{i+1})$ is an edge for all $0 \leq i \leq l^k - 1$ and denote the vertices of the obtained digraph as $v_i^a$, where $0 \leq a \leq m_k - 1$. The cycle is then formed by edges $(v_0^0, v_0^1), \ldots, (v_0^{l^k-2}, v_0^{l^k-1})$ and an edge $(v_0^{l^k-1}, v_1^0)$, and so on.

The obtained digraph $D_k$ has $n_k$ vertices and in-degree $d_k = 2^k$, and hence $f(n_k) \geq d_k$. Furthermore, it can be easily shown that this digraph has girth $l$ and satisfies $\frac{g_{\text{linear}}(D_k, s)}{n_k} \geq \frac{g_{\text{linear}}(C_l^k, s)}{l^k} \geq 1 - \left(\frac{l-1}{l}\right)^k$ by Example 11, which tends to 1. ∎

Theorem 3 implies that there exist network coding instances with a relatively small number of intermediate nodes, a relatively small number of edges coming in or out each node, and an arbitrarily long path between each source and its corresponding sink. These instances are linearly solvable over any alphabet, and the operation at each node is known.

## VI. Conclusion and open problems

In this paper, we proved that the problem of deciding whether a network coding instance was solvable reduced to a problem on the independence number of a related undirected graph, referred to as the guessing graph. Although we have derived bounds on this independence number, how to efficiently compute it remains an open problem. A brute force approach would be computationally infeasible, as the maximum independent set problem is NP-hard. Also, algorithms for the maximum independent set problem on general graphs are inappropriate, for the size of the guessing graph grows exponentially with the number of nodes in the original network coding instance. However, the guessing graph has many symmetries (its structure is fixed by the original instance), hence specific algorithms could be devised to bound or compute its independence number. The relationships between this problem and coding theory is of peculiar interest. In particular, we exhibited classes of network coding instances for which the maximum independent set of the guessing graph is given by cyclic codes.

The second contribution of our paper is the design of a family of digraphs for which the ratio between the guessing number and the number of vertices tends to one, although they have a large girth and are sparse. This family of digraphs yields a family of solvable network coding instances, for which binary linear operations are sufficient. Although we gave necessary and sufficient conditions on the sparsity of the graph in terms of edges, the maximum speed of convergence to one of the ratio remains unknown. Similarly, the relation between the guessing number and the girth seems an interesting problem for network design.

## VII. Acknowledgments

We would like to thank Antonia Katzouraki, the Associate Editor Massimo Franceschetti, and the anonymous Reviewers for their valuable comments.